# A Glimpse of the First Eight Months of the COVID-19 Literature on Microsoft Academic Graph: Themes, Citation Contexts, and Uncertainties

Chaomei Chen
College of Computing and Informatics
Drexel University, 3141 Chestnut Street, Philadelphia, PA 19104, USA
Email: cc345@drexel.edu

**Abstract**
As scientists worldwide search for answers to the overwhelmingly unknown behind the deadly pandemic, the literature concerning COVID-19 has been growing exponentially. Keeping abreast of the body of literature at such a rapidly advancing pace poses significant challenges not only to active researchers but also to the society as a whole. Although numerous data resources have been made openly available, the analytic and synthetic process that is essential in effectively navigating through the vast amount of information with heightened levels of uncertainty remains a significant bottleneck. We introduce a generic method that facilitates the data collection and sense-making process when dealing with a rapidly growing landscape of a research domain such as COVID-19 at multiple levels of granularity. The method integrates the analysis of structural and temporal patterns in scholarly publications with the delineation of thematic concentrations and the types of uncertainties that may offer additional insights into the complexity of the unknown. We demonstrate the application of the method in a study of the COVID-19 literature.

**Introduction**
The COVID-19 pandemic has impacted so many people's everyday life worldwide and it is still threatening our society as a whole. The COVID-19 pandemic is unprecedented in several ways that make it particularly challenging and threatening: it is still largely unknown of its origin and transmission routes; there may be months or even longer before COVID-19 vaccines can be expected to be a powerful line of defense; the prolonged pandemic continues to pose social, economic, and political challenges to businesses and entire industries such as airlines and international travels as well as schools and many other areas.

Scientists and researchers have actively responded to the urgency and severity of the pandemic. Publications relevant to COVID-19 have increased rapidly across disciplines since the beginning of the year 2020. Several institutions and corporate organizations have contributed openly accessible datasets of COVID-19 publications, notably including the CORD-19 dataset and the Lens.

The CORD-19 dataset, the COVID-19 Open Research Dataset, covers the scholarly literature of COVID-19, SARS-CoV-2, and the coronavirus group[1]. Its initial release contained over 29,000 articles, over 13,000 of which contained full text. The CORD-19 dataset has been updated regularly. By the beginning of September 2020, the CORD-19 dataset reached 130,000 articles. It has been studied by many researchers, especially from communities that are well-equipped to analyze and model text documents, including data science and AI for example. A scientometric researcher, however, may find the CORD-19 dataset and other similar datasets inadequate, for example, due to the lack of cited references as part of the dataset, missing elements such as abstracts, or other data coverage or quality issues.

---

[1] https://www.whitehouse.gov/briefings-statements/call-action-tech-community-new-machine-readable-covid-19-dataset/

Several widely used analytic methods in the field of scientometrics rely on citations in scholarly publications to derive indicators and metrics of underlying research and its impact, for example, including the h-index and the g-index. Eugene Garfield pioneered the notion of Science Citation Index in the 1950s, which subsequently formed the basis for citation analysis and quantitative studies of science in general (Garfield 1955). Typically, authors of a scholarly publication refer to, i.e. cite, previously published works in their narratives. There are many studies of how and why authors ought to cite their references appropriately and how and why they often failed to do so (Greenberg 2009). The Web of Science, Scopus, Dimensions, the Lens, and Microsoft Academic (MA) are among the most widely used multi-disciplinary bibliographic databases (Visser, Eck et al. 2020), although they differ considerably in terms of the breadth and depth of their coverage, data quality, and accessibility. Document Co-Citation Analysis (DCA) (Small 1973) and Author Co-Citation Analysis (ACA) (White and McCain 1998) are network-based analytic methods for studying scientific literature. Networks in such studies are constructed based on co-occurrences of a pair of entities such as cited references or cited authors. The resultant networks then lend themselves to a wide variety of network analysis, modeling, and visualization techniques. Although networks can be derived from co-occurrences of words and phrases in text, references may play valuable roles at a higher level of abstraction, especially in what is known as concept symbols (Small 1978, Small 1986). A major advantage of bibliographic databases such as the Web of Science and Scopus is the provision of references cited by a scholarly publication. These databases are conveniently used by researchers to conduct citation analysis and other citation-based studies. In contrast, PubMed, a well-known resource of the literature of biomedicine, does not provide references of an article as part of its metadata. CORD-19 provides a similar depth of coverage as PubMed, i.e. the references cited by an article are not readily available from the dataset, which prevent researchers from conducting citation-based network analyses.

Microsoft Academic (MA) is a major source behind the construction and dissemination of CORD-19. Microsoft Academic Graph (MAG) organizes entities and relations of scholarly publications as a graph and allows flexible ways to retrieve bibliographic data, including references cited by an article (Wang, Shen et al. 2019). Furthermore, instances of a citation of a reference, often known as citances (Nakov, Schwartz et al. 2004), are also retrievable from MAG in terms of citation contexts. Citation contexts of a reference consist of sentences in which the reference is cited along with surrounding sentences to provide a meaningful context.

In the rest of the article, we will introduce a generic and reproducible method based on MA for visual analytic studies of the COVID-19 literature. In fact, the method is applicable to the literature of other topics. This method contributes to the study of a research field as follows: 1) enable anyone who is interested to construct their own CORD-19-like dataset with an enriched inclusion of cited references and citation contexts; 2) eliminate the barrier to conduct scientometric studies with the citation enriched data; 3) expand the current visual analytic studies of research literature further with the inclusion of citation contexts and enhance network-based analyses with deeper insights identified at a finer level of granularity; 4) enable analysts to interactively exploring uncertainties associated with the citations of a reference; and 5) conduct these analyses in an integrated visual analytic environment – an enhanced implementation of a widely used science mapping tool – CiteSpace (Chen 2004, Chen 2006, Chen, Ibekwe‐SanJuan et al. 2010, Chen 2017).

**CiteSpace**
CiteSpace is designed for conducting a visual analytic study of the scholarly literature of a research field, a research area, or a discipline, collectively known as a knowledge domain (Chen 2004, Chen 2006). CiteSpace constructs a series of networks of underlying entities and their relations derived from a representative dataset of the corresponding knowledge domain. Structural patterns and trends are combined with temporal patterns and indicators to inform analysts significant developments of a research field in question. A typical workflow divides a synthesized network into distinct clusters of entities such as cited references. The analyst would focus on the meaning of individual clusters and their interrelationships so as to uncover critical insights from high-levels of aggregation to lower levels. For example, applications of CiteSpace typically aim to address the following questions:
- What are the major thematic concentrations in the field of research in terms of clusters of co-occurring references?
- How are neighboring clusters connected? Which articles serve as bridges between distinct clusters?
- Which articles are the most representative members of a particular cluster?
- What would be the most accurate word or phrase to sum up the role of a cluster?

Technically, CiteSpace decomposes a network into distinct clusters. Each cluster consists of a set of references that are frequently cited together by the same citing articles. In other words, the member references of a cluster are more often cited within the same article than references from different cluster. The quality of the decomposition is measured in terms of network modularity, the silhouette score of each cluster, and a cluster-size weighted silhouette mean. The closer these values are to 1.0, then higher the overall clarity of the configuration.

Conceptually, the position of a reference in the underlying network representation can be used to guide our navigation. According to the Structural Hole Theory (Burt 2004), nodes located at structural holes tend to be associated with creativity, originality, and boundary-spanning. In CiteSpace, structural holes are highlighted as nodes with high betweenness centrality scores. Landmarks provide useful guidance for navigation. Highly cited articles are depicted as large tree-rings of citations per year. Articles with sharp increases in citations they received experience citation bursts, which indicate their particularly noteworthiness because of their extraordinary attraction of attention.

CiteSpace supports Structural Variation Analysis (SVA), which is a predictive analytic process that can be used to identify newly published articles with transformative potentials (Chen 2012). Transformative potentials are measured in terms of significant structural variations induced by newly published articles. At a given time, a newly published article contributes to the literature by connecting previously disjoint bodies of thematic concentrations. Such connections may lead to transformative changes to the underlying knowledge structure. Transformative changes are realized if the research community follows and consolidates the promising paths forged by the innovative connections. On the other hand, not all such connections lead to changes that are responded by the research community, thus they are considered as the potential at the point of recognition other than as actual transformative changes.

The role of uncertainties in representing scientific knowledge is proposed in our recent work to capture the state of the art of a research field as well as to explain what drives the dynamics of underlying research activities (Chen and Song 2017, Chen, Song et al. 2018). We proposed a framework of uncertainties with three major types of uncertainties that are measurable at the linguistic level, namely, epistemic uncertainty (E), hedging (H), and transitional (T). Epistemic uncertainty is measured based on the presence of uncertainty cue words that indicate the unknown, incomplete, controversial, and contradictory aspects of a topic. Hedging is measured in terms of hedging words such as may, suggest, and imply. Hedging words are essential linguistic devices to render the certainty or, rather the uncertainty, of a statement more appropriately. Recently, there is a growing interest in the connection between hedging and citations (Small 2018, Small, Boyack et al. 2019). Transitional uncertainties are measured by the presence of transitional words such as however and nonetheless, which often signify a change in an argument or the possibility of multiple alternative interpretations of a situation. These uncertainties are measured in terms of information entropy. Using information entropy has several advantages over other metrics. First, information entropy has been widely used to represent uncertainties. Second, measuring uncertainties at different levels of aggregation follows naturally with information entropy. For example, the uncertainty associated with an article can be defined as the sum of all the uncertainties associated with the text passages of the article, such as its citation contexts. Similarly, the uncertainty of a cluster of cited references is naturally the sum of the uncertainties of all its member references.

Currently, citation contexts are not available from the most commonly used bibliographic databases, such as the Web of Science, Scopus, Dimensions, and the Lens, except Microsoft Academic Services (MAS)(Sinha, Shen et al. 2015, Wang, Shen et al. 2019).

**Microsoft Academic Services**
Microsoft Academic Services (MAS) provides two ways to access the Microsoft Academic Graph (MAG): through API to access the MAG hosted by MAS or maintain a copy of MAG of your own. The potential and limitations of Microsoft Academic (MA) for citation analysis have been studied and compared with other commonly used resources (Hug and Brandle 2017, Hug, Ochsner et al. 2017, Thelwall 2017, Thelwall 2018, Visser, Eck et al. 2020). More recent reviews of MAS are presented in (Wang, Shen et al. 2019). An interesting study of the core concepts of classic philosophical books made a good use of the MAG (Bornmann, Wray et al. 2020).

Each article in MAG has a MAG ID, for example, 2160441315 for (Chen 2004) and 2150999626 for (Chen 2006). One may retrieve publications from MAG based on their MAG IDs, DOIs, and fields of study. It is also possible to retrieve articles that cite a set of references in their MAG IDs, thus this function enables us to enrich an article's metadata with its references. For each returned article, citation contexts provide a unique advantage of MAS over other major bibliographic databases. The integration of visual analytics of citation contexts with other citation-based analytic processes is a significant contribution of our new method.

To demonstrate the application of the integrated and enhanced method, we focus on the COVID-19 literature. Instead of using the currently available CORD-19 dataset due to its lack of cited references, we construct a dataset of the COVID-19 literature directly from MAS. A major advantage of this approach is its flexibility and extensibility. Anyone who is interested would be able to perform the construction with the new version of CiteSpace on a subject of their own interest. In particular, it is possible to further extend CiteSpace to support Cascading Citation Expansion (Chen and Song 2019), which is so far only feasible in CiteSpace with Dimensions' API. At the time of writing, the Lens API currently does not support retrieving citing articles of a given reference, although the Lens ScholarlyWorks API is very powerful in other aspects and it is supported in CiteSpace.

We constructed the MA-based COVID-19 dataset based on fields of study. The following query means a qualified record should match with at least one of the fields of study.

```
expr=Or
    (Composite(F.FN=='coronavirus disease 2019'),
    Composite(F.FN=='severe acute respiratory syndrome coronavirus 2'),
    Composite(F.FN=='2019 20 coronavirus outbreak'),
    Composite(F.FN=='covid-19'),
    Composite(F.FN=='covid19')
)
```

The dataset constructed on September 5, 2020 contains a total of 79,476 records and 7,693 of them have corresponding citation contexts. This dataset represents a snapshot of the COVID-19 literature published in the first eight months of 2020 on MAG. 77,897 records have been cited at least once. These records are stored in a format that is a superset of the Web of Science field-delimited data format. The citation contexts are stored in a separate file to maximize the interoperability. The integrated analytic functions involving citation contexts use a database on a local host. Users who do not install such databases can still use CiteSpace to process the dataset retrieved from MAS as if they were from the Web of Science.

The MAS-enriched COVID-19 dataset can be studied at three levels. First, the dataset contains cited references the same way as bibliographic records of the Web of Science origin. All the visual analytic functions in CiteSpace that can be applied to the Web of Science data now can be applied to the MAS-enriched dataset. We will illustrate a visualization of the co-citation network with monthly time-sliced intervals and an overview of its major thematic concentrations (i.e. clusters of co-cited references). Second, the integrative analysis of citation contexts along with Level-1 analyses, which are visual analytic tasks typically used with tools such as CiteSpace. Citation contexts serve the source of information from which three types of uncertainties are measured. One reference can be cited by the same citing article in multiple citation contexts as well as by multiple citing articles in an even larger number of citation contexts (Ding, Liu et al. 2013, Hu, Chen et al. 2013). Making sense of key themes in the potentially diverse and extensive narratives is a time consuming and cognitively demanding task. Uncertainty scores provide an extra layer of organizing the information and highlight information that is potentially valuable to the sense-making process. Finally, at the third level, we demonstrate the application of the SVA procedure to the dataset and identify a set of newly published articles that are too young to standout in terms of their citation counts but start to show signs of transformative potential.

**Data**

To demonstrate the method, we constructed a dataset of the COVID-19 literature from the MAG based on matching fields of study. The resultant dataset contains publications as earlier as 2014. Interestingly, the dataset also contains publications from the future, all the way from next month to next year. Table 1 summarizes the dataset. Our subsequent discussions will focus on the subset of the first eight months in 2020.

*Table 1. A self-constructed dataset of the COVID-19 literature on Microsoft Academic Graph as of September 5, 2020.*

| Time of Publication | Unique Citing Articles | Unique References | Citation Contexts | Unique Contexts |
|---|---|---|---|---|
| 2014-SEP | 1 | 14 | 38 | 23 |
| 2018-FEB | 1 | 57 | 92 | 59 |
| 2018-NOV | 1 | 18 | 32 | 26 |
| 2019-JAN | 2 | 18 | 22 | 14 |
| 2019-JUN | 1 | 5 | 15 | 15 |
| 2019-MAR | 1 | 81 | 172 | 73 |
| 2019-MAY | 1 | 246 | 443 | 221 |
| 2019-SEP | 1 | 10 | 17 | 17 |
| 2019-DEC | 4 | 102 | 157 | 96 |
| 2020-JAN | 241 | 1939 | 3476 | 2529 |
| 2020-FEB | 241 | 1601 | 3287 | 2251 |
| 2020-MAR | 997 | 5704 | 13206 | 9596 |
| 2020-APR | 2765 | 15478 | 36599 | 26507 |
| 2020-MAY | 1657 | 11227 | 23131 | 17089 |
| 2020-JUN | 756 | 6445 | 12144 | 9260 |
| 2020-JUL | 687 | 6218 | 11837 | 8640 |
| 2020-AUG | 300 | 3462 | 5845 | 4304 |
| 2020-SEP | 18 | 230 | 371 | 292 |
| 2020-OCT | 8 | 193 | 287 | 231 |
| 2020-NOV | 5 | 76 | 138 | 107 |
| 2020-DEC | 3 | 17 | 24 | 22 |
| 2021-JAN | 2 | 20 | 27 | 23 |

Table 2 lists the number of articles found on major bibliographic databases on September 5, 2020 with the identical or closely matched queries to the query used for this study. For example, the following query was used to search in the titles only in Dimensions DSL. For full text search, replace the scope title_only with full_data.

```
search publications in title_only for "\"coronavirus disease 2019\" OR
\"severe acute respiratory syndrome coronavirus 2\" OR \"2019 20
coronavirus outbreak\" OR \"covid-19\" OR covid19" return publications
[id]
```

A search on Google Scholar used the following query:
```
'coronavirus disease 2019' OR 'severe acute respiratory syndrome
coronavirus 2' OR '2019 20 coronavirus outbreak' OR 'covid-19' OR
'covid19'
```

The search on MAG used the following query:
```
expr=Or(Composite(F.FN=='coronavirus disease 2019'),
Composite(F.FN=='severe acute respiratory syndrome coronavirus 2'),
Composite(F.FN=='2019 20 coronavirus outbreak'),
Composite(F.FN=='covid-19'), Composite(F.FN=='covid19'))
```

The CORD-19 dataset is the most comprehensive one with 130,000 articles. Full text searches on Dimensions and the Lens returned over 100,000 articles. Searches in the metadata alone of course returned fewer articles than full text searches. We used fields of study to construct our demonstrative dataset to reduce the complexity of the query formation. More sophisticated strategies such as Cascading Citation Expansion can be used to optimize the overall quality of the data collection step (Chen and Song 2019). A topic search in the Web of Science returned the lowest number of articles in this group. In terms of the percentage of the CORD-19 dataset, the

Web of Science represents about 23% of its volume, MAG over 62%, and full text searchers on Dimensions and Lens over 75%.

*Table 2. A simple comparison between different data sources on the search query used for this study as of September 5, 2020.*

| Data Source | Articles | % of CORD-19 | Search Strategy |
|---:|---:|---:|---|
| CORD-19[2] | 130,000 | 100.00 | combined |
| Dimensions[3] | 124,131 | 95.49 | full text |
| Lens[4] | 100,759 | 77.51 | full text |
| Lens | 98,839 | 76.03 | title, abstract, keyword |
| Dimensions | 90,747 | 69.81 | title, abstract, keyword |
| Lens | 83,048 | 63.88 | title |
| MAG[5] | 80,676 | 62.06 | fields of study |
| Dimensions | 75,435 | 58.03 | title |
| Google Scholar[6] | 73,700 | 56.69 | title, abstract, full text |
| Web of Science[7] | 29,858 | 22.97 | topic search |

**Overview**

The first question regarding the literature of a topic that is full of uncertainty is: what are the major components of the COVID-19 literature? What aspects of the subject are predominant in the first eight months of the pandemic?

Figure 1 shows an overview of the underlying network of references that are often cited together. The nodes are references cited by citing articles, which are records from the dataset retrieved from MAG, whereas links between them represent the strengths of their co-occurrences. Groupings of references, i.e. clusters, emerge as some references are co-cited more often than others. These clusters represent concentrations of themes, although their degrees of concentration may vary widely across different clusters. Each cluster is assigned an automatically generated cluster label, for example, #3 spike protein and #5 pregnant women. Clusters are numbered from #0 onwards. Clusters are depicted in different colors, starting with the red for the largest cluster #0 and follow the rainbow colormap. We are of course aware that the rainbow colormap may be harmful if used inappropriately but in this case as long as we can differentiate one cluster from another, it would be sufficient.

The largest cluster #0 wuhan china is shown in red near the center of the network. The second largest cluster #1 renin angiotensin aldosterone system is located near the top. Clusters #2 systematic review meta analysis, #3 spike protein, #5 pregnant women can be found on the left. On the right half of the graph, there are clusters #4 dental practice, #6 chest [CT], and #12 mental health.

The configuration of the parameters for this network is shown in the upper left corner. Users may adjust them accordingly. The provision of citation contexts provides another option to enhance the conventional Document Co-Citation Analysis (DCA), which typically treats each reference with an equal weight for a citing article because no further data beyond the reference list is available. In this study, we utilize the additional information accessible from the corresponding citation contexts such that references frequently cited in the article can be retained, whereas references cited below average can be discarded. Users may control this option.

---

[2] https://www.semanticscholar.org/cord19
[3] https://app.dimensions.ai/discover/publication
[4] https://www.lens.org/
[5] https://docs.microsoft.com/en-us/academic-services/
[6] https://scholar.google.com/schhp?hl=en
[7] https://clarivate.com/webofsciencegroup/solutions/web-of-science/

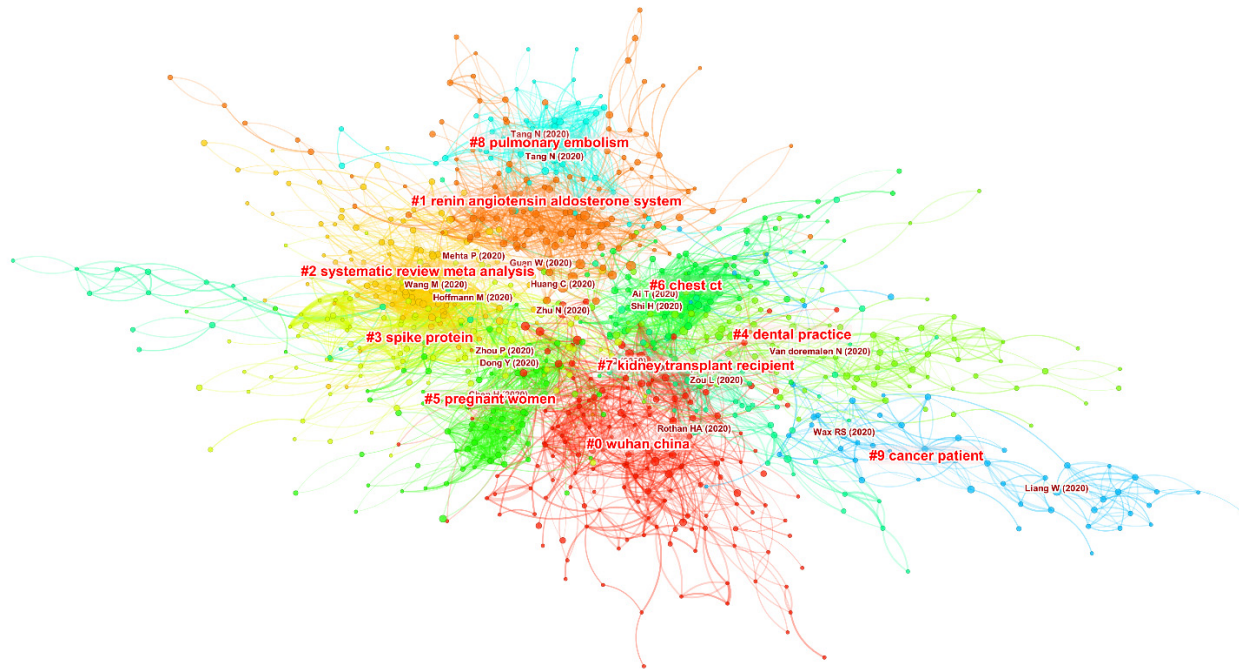

*Figure 1. An overview of 1,330 top-cited articles in the COVID-19 literature of 77,897 articles.*

**Concept Trees**

The MAG-based COVID-19 dataset contains 111,360 citation contexts made by 7,693 citing articles, involving 39,995 references. Although this represents only 9.88% of the 77,897 citing articles that have at least one citation in the dataset, the available citation contexts provide enough information for us to assess the feasibility of the integrative approach.

The overview in Figure 1 gives us some general idea about the most active areas of the literature. To understand what each cluster involves, one option is to characterize the thematic structure of the cluster in a hierarchical representation of its component concepts (Palla, Tibely et al. 2015, Balogh, Zagyva et al. 2019). Figure 2 shows a concept tree of Cluster #5 pregnant women. The nodes of the tree are concepts, which are noun phrases extracted from citation contexts of the member references in this cluster. Major branches of the tree highlight the key themes or concerns of the cluster. For example, vertical transmission is shown as the root of the tree, which is the primary concern of this cluster. Users may interactively inspect the citation contexts of each concept, as shown in the right window in Figure 2. The citing paper's MAG ID is shown as a hyperlink. Following the hyperlink will take the users to the record displayed on the Microsoft Academic website.

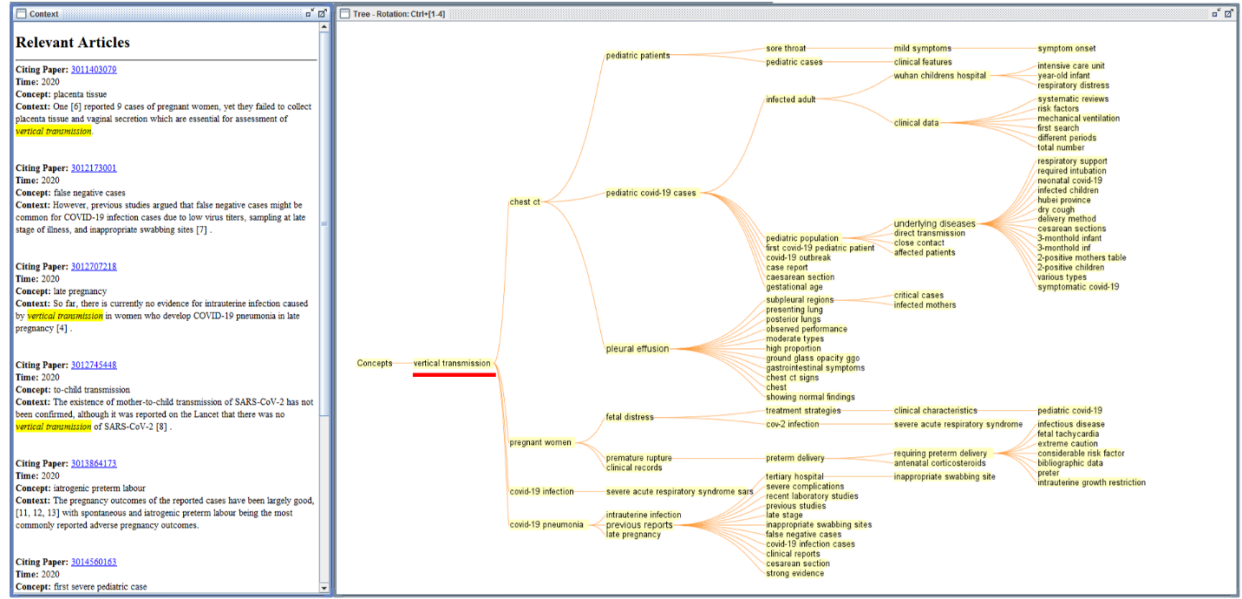

*Figure 2. Making sense of a cluster (#5 pregnant women). The list of citation contexts shown in the left window corresponds to the currently mouse-over event on the concept of vertical transmission.*

Concept trees can be used to highlight the structure of a cluster, a reference, or a word or a phrase. Figure 3, for example, is generated from a single reference that has the highest epistemic uncertainty score. The concept tree reveals that incubation period and mean incubation period are major themes emerged from citation contexts made by subsequently published articles. Interactive inspections of the concept of mean incubation period reveal specific days mentioned in various citation contexts shown in the left window. For example, the mean incubation period of 5.2 days appears frequently in these contexts. This function can facilitate traditional systematic reviews and meta analyses as one can efficiently collect values of a specific variable from a large number of publications.

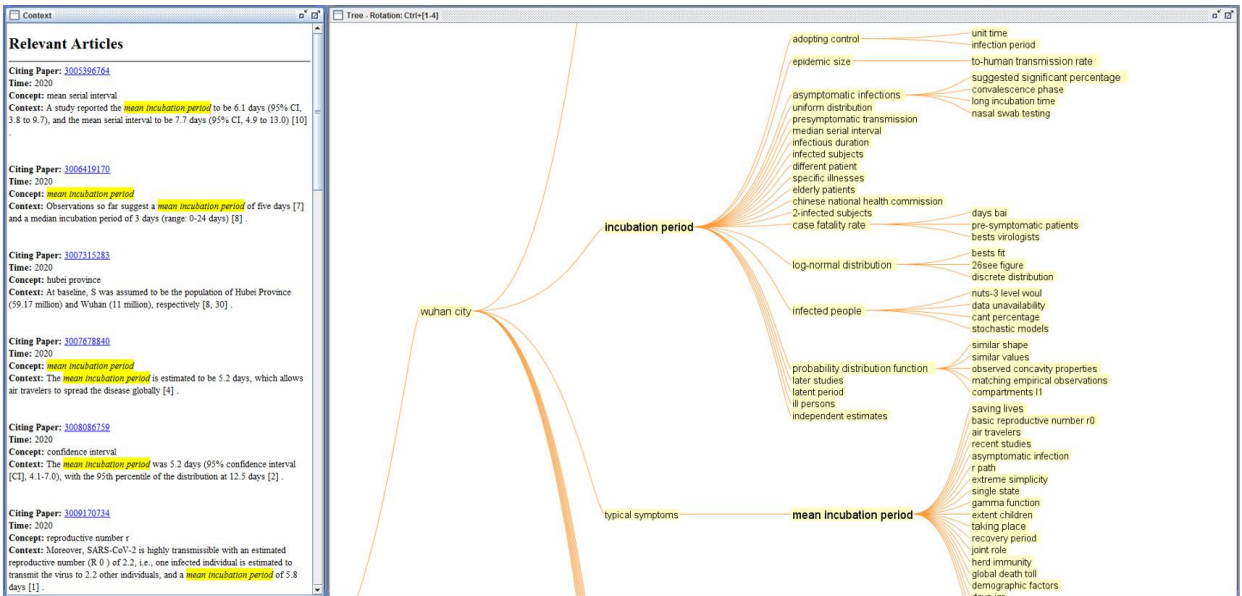

*Figure 3. Making sense of major themes of citations to a specific reference (Li, Q., 2020).*

**Uncertainties**

Each citation context consists of a few sentences in which a particular reference is cited by a citing article. The uncertainties of these sentences can be measured in terms of the presence of uncertainty cue words and how often these cue words appear in a global bibliographic database such as the entire collections of over 200 million articles on Dimensions. Semantically equivalent cue words are learned from initially hand-picked cue words as seeds (Chen, Song et al. 2018). Epistemic uncertainties are designed to capture uncertainties in our knowledge, which can be manifested by the appearance of cue words such as controversial, contradictory, inconsistent, unknown, and uncertain.

While the uncertainties of citation contexts can be identified independent of the underlying topic as shown in the above example, a more valuable option is the combine the level of uncertainty with specific rhetorical cues or specific concepts. For example, we may want to see how epistemic uncertainties are associated with concrete concepts such as social distancing, face masks, and international travels. As shown in Table 3, a combination of epistemic uncertainty cues and rhetorical words on conclusions reveals additional insights into the challenges we are facing.

*Table 3. Epistemic uncertainties of citation contexts containing specific rhetorical words on conclusions. Uncertainty cue words are highlighted in yellow, whereas rhetorical words are highlighted in green.*

| Citing Article→Cited Reference | Epistemic Uncertainty | Citation Context: Uncertainty: uncertain/conflicting/contradict/inconsistent Rhetorical: conclusion/conclude |
|---|---|---|
| 3037877512→3018691224 | 0.0314 | rly complex and counter-intuitive due to the uncertainty in the transmission mechanisms, possible seasonal variation in both susceptibility and transmission, and their variation within subpopulations[7]. The media has given extensive coverage to analyses and forecasts using COVID-19 models, with increased attention to cases of conflicting conclusions, giving the impression that epidemiological models |
| 3079224143→3013360115 | 0.0314 | [35] conclude that not only the COVID-19 numbers will grow but also uncertainty about forecasts will also grow. |
| 3020670761→3010344953 | 0.0314 | (14) Finally, a recent systematic review of the literature concluded that while further research on this topic was required, the limited data available, albeit not adjusted for other factors potentially impacting on outcome, concluded that smoking was most likely associated with negative progression and outcomes of SARS-CoV-2,(15) while a second meta-analysis concluded that active smoking was not associated with severity of SARS-CoV-2.(16) Nevertheless, despite this current uncertainty many organisations, such as the World Health Organization and the National Institute of Drug Abuse, recommend smoking cessation strategies not only to alleviate the harm caused by smoking, in general, but also because smoking cessation may potentially lessen the risks of SARS-CoV-2 infection. |
| 3023144169→2969352266 | 0.0237 | It can be concluded that the results of previous studies on the immune system have been inconsistent. |
| 3009935283→2811210701 | 0.0205 | 21 The conflicting conclusions from our two scenarios, driven largely by the differences in the extent of presymptomatic transmission, highlight the urgent need for more data to clarify key epidemiological parameters of COVID-19, particularly the serial interval and the extent of presymptomatic transmission, in order to inform response efforts. |
| 3021685303→280 | 0.0100 | Moreover, the contradictory findings about smoking in the literature, with |

| 2058961 | | | studies published before the COVID-19 pandemic reporting that smoking and nicotine down-regulate ACE2 [3, 4] and other studies published during the pandemic reporting that they up-regulate ACE2 [5] [6] [7] , do not allow for solid conclusions regarding the effects of nicotine or smoking on ACE2. |
|---|---|---|---|
| 3007114958→3002108456 | | 0.0014 | Compared with the results of the two studies on Wuhan cases by Chen et al 18 and Huang et al 19 , we found that the gender proportion was equal in the 80 patients we included, contradicting to the conclusion that men were more susceptible than women. |

Figure 4 shows a screenshot of the Node Details window in CiteSpace for the article (Li, Guan et al. 2020), which has the largest epistemic uncertainty score in the dataset. The screenshot shows a list of citation contexts of the reference in chronological order. The length of an orange bar depicts the epistemic uncertainty score (E). Major uncertainty cue words contributing to the uncertainty score are highlighted in red, including unknown, uncertainty, and controversial. The visual representation reduces the cognitive burdens for us to manually sift through numerous citation contexts across different articles. It makes it easier for us to concentrate on the key information. We can quickly scan citation contexts and focus on ones with high uncertainty scores. It is out belief that uncertainties can provide valuable insights into the heart of research (Chen and Song 2017).

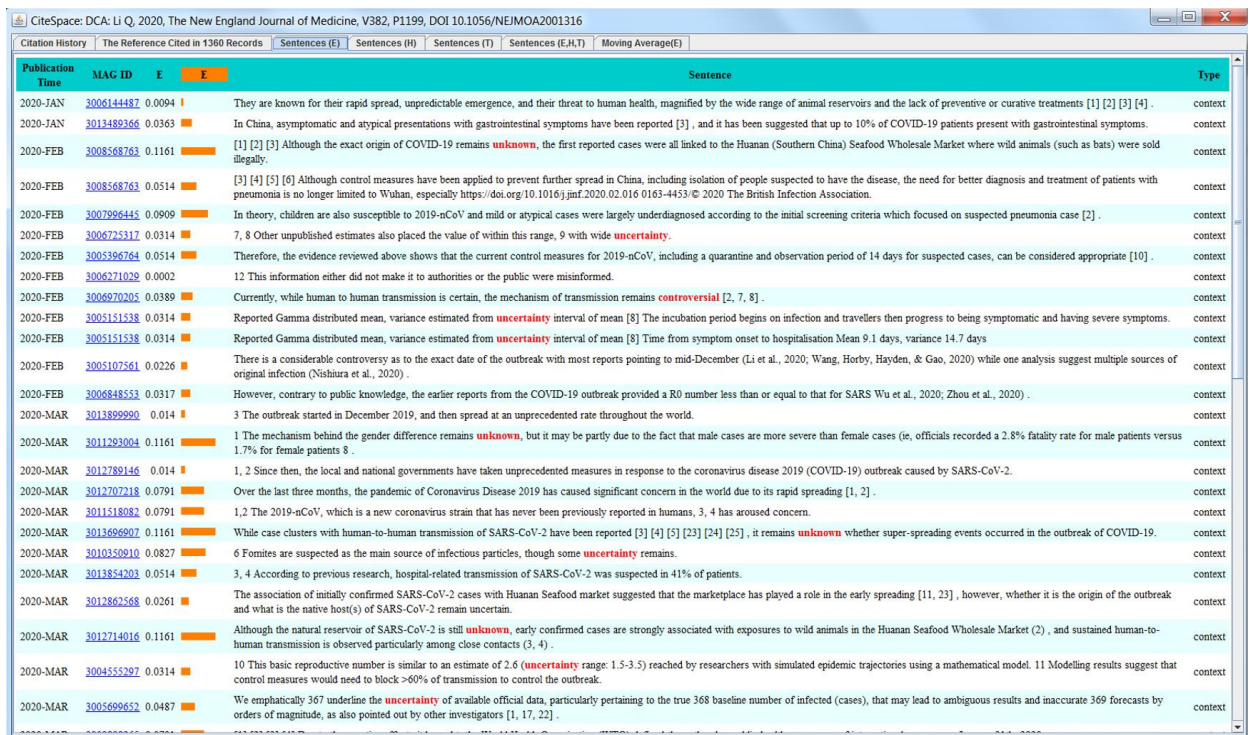

*Figure 4. Uncertainties of contexts in which (Li, Guan et al. 2020) has been cited.*

Figure 5 shows a concept tree of a phrase, vaccine. The concept tree on the left is the entirety of the hierarchical representation of the concepts co-occurring with the words vaccine or vaccination. Similarly, one can interactively inspect the concept tree and the underlying citation contexts.

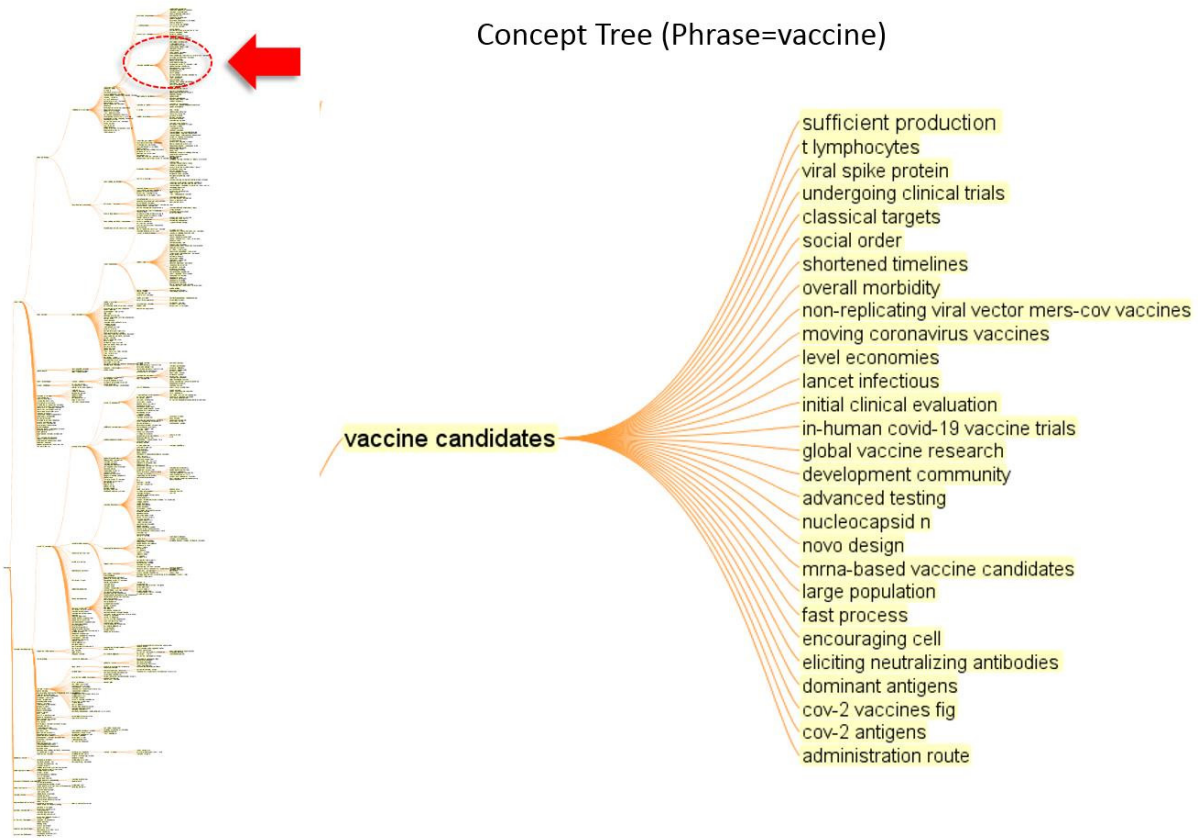

*Figure 5. Concept tree of a phrase: vaccine.*

The examples shown above illustrate the application of the method on datasets retrieved from MAG. The following examples demonstrate Structural Variation Analysis (SVA), which is a predictive analytic method to identify newly published articles with transformative potentials (Chen 2012). In other words, SVA aims to identify new articles that may have profound impacts on the underlying literature in the future.

**Structural Variation Analysis (SVA)**
The structural variation analysis is built on the basis of Structural Hole Theory, which was originally proposed for social networks (Burt 2004, Chen 2012). According to the Structural Hole Theory, people's positions in their social network or a network of their co-workers may be associated with social capitals, which may be in turn translated into competitive edge. Furthermore, positions that play a significant role in connecting different parts of the network tend to have a greater potential than other positions. From an information flow point of view, individuals who are in such positions are exposed to different ideas, perspectives, and opinions, which make them more open-minded and creative. We extended this insight to networks of scholarly publications. References located at such positions, i.e. connections are limited or completely missing, have the potential to make a profound impact on the global structure of the underlying research field. The SVA procedure looks for newly added connections that may alter the global structure or have the potential to do so. In essence, each newly published article is compared against a baseline network formed by the literature up to the publication of the article in question. Co-occurring links made by the article are examined in the baseline network to determine whether a specific link is novel between clusters or incremental within clusters or anything else in between. Transformative potential scores are given to articles that make novel links between distinct clusters.

Figure 6 shows the overview of the network we will work with for SVA. The current implementation of SVA is computationally expensive; therefore, we used a smaller network than the one we showed earlier for the demonstration. The overview shows a similar set of clusters with slight shifts of focus. The largest cluster is #0 spike protein. The second largest one is #1 transmission route. The three articles labeled in black background are the three articles with the highest epistemic uncertainty scores, including the article (Li, Guan et al. 2020), which we have seen earlier in the study.

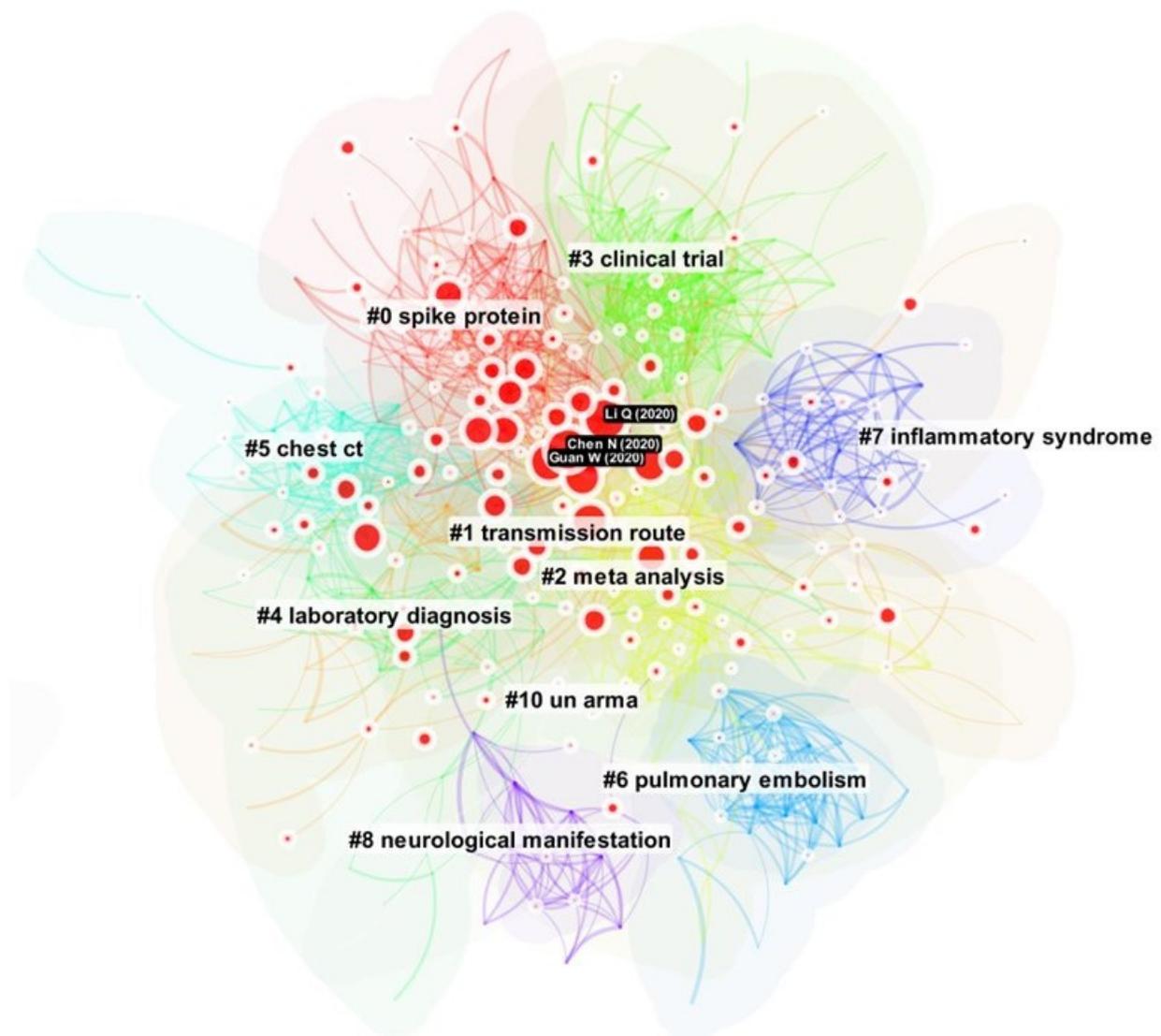

*Figure 6. An overview of a smaller network to illustrate SVA. The size of a disc in red depicts the epistemic uncertainty (E). The largest three discs are labeled in black background.*

Figure 7 shows the same set of clusters in a more separated arrangement. The separation reveals that the largest cluster #0 spike protein contains the most nodes with significant levels of epistemic uncertainty. One of the three most 'uncertain' nodes, (Li, Guan et al. 2020), is in this cluster, whereas the other two belong to cluster #2 meta analysis. In contrast, the overall uncertainty of cluster #6 is very low. The distribution of epistemic uncertainty at the cluster level requires further investigation. One possible direction would be along the hypothesis that a high epistemic uncertainty indicates a sophisticated level of articulation in scholarly communication. Another possible direction would be that a high concentration of epistemic uncertainty articles may indicate the origin of a field of research when researchers conceptualize the fundamental research needs.

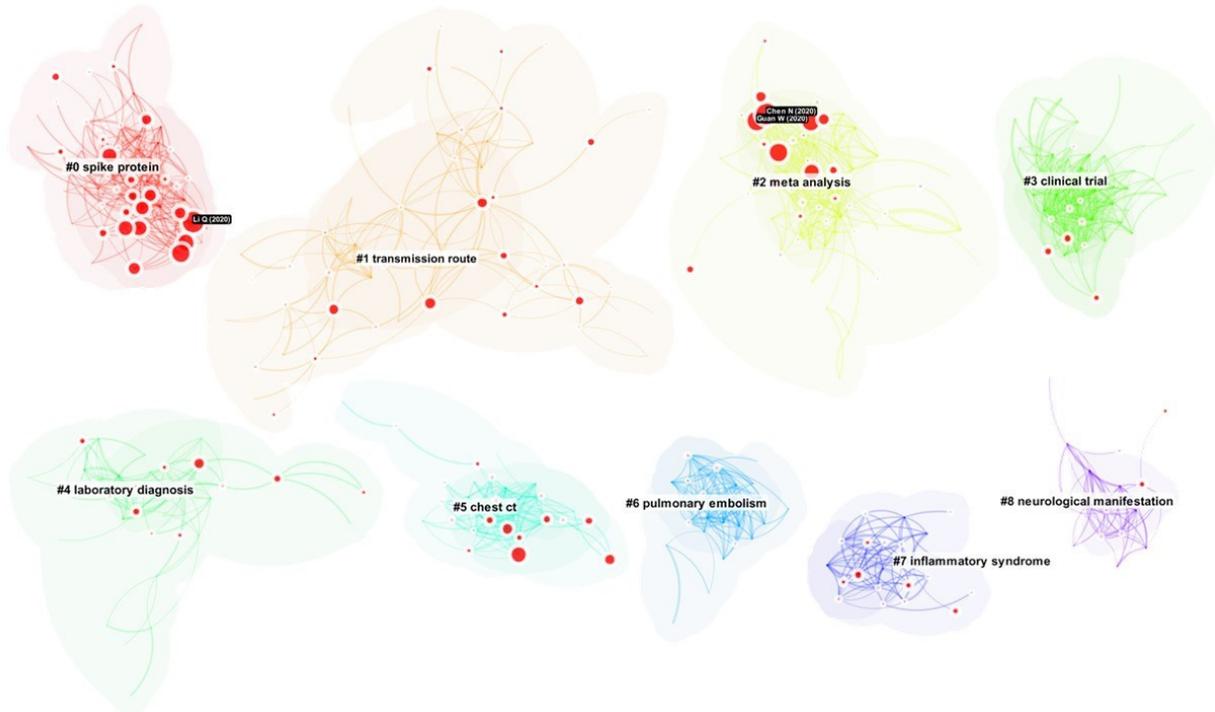

*Figure 7.The distribution of citation contexts with uncertainties is uneven. The most of uncertainties are in clusters 0, 2, and 5.*

Figure 8 shows an overlay of cluster #0 spike protein over the overall network. Such overlays highlight the scope of a cluster and inter-cluster relations. The visualization also suggests that the majority of the epistemic uncertainty appear to concentrate on the central area of the network.

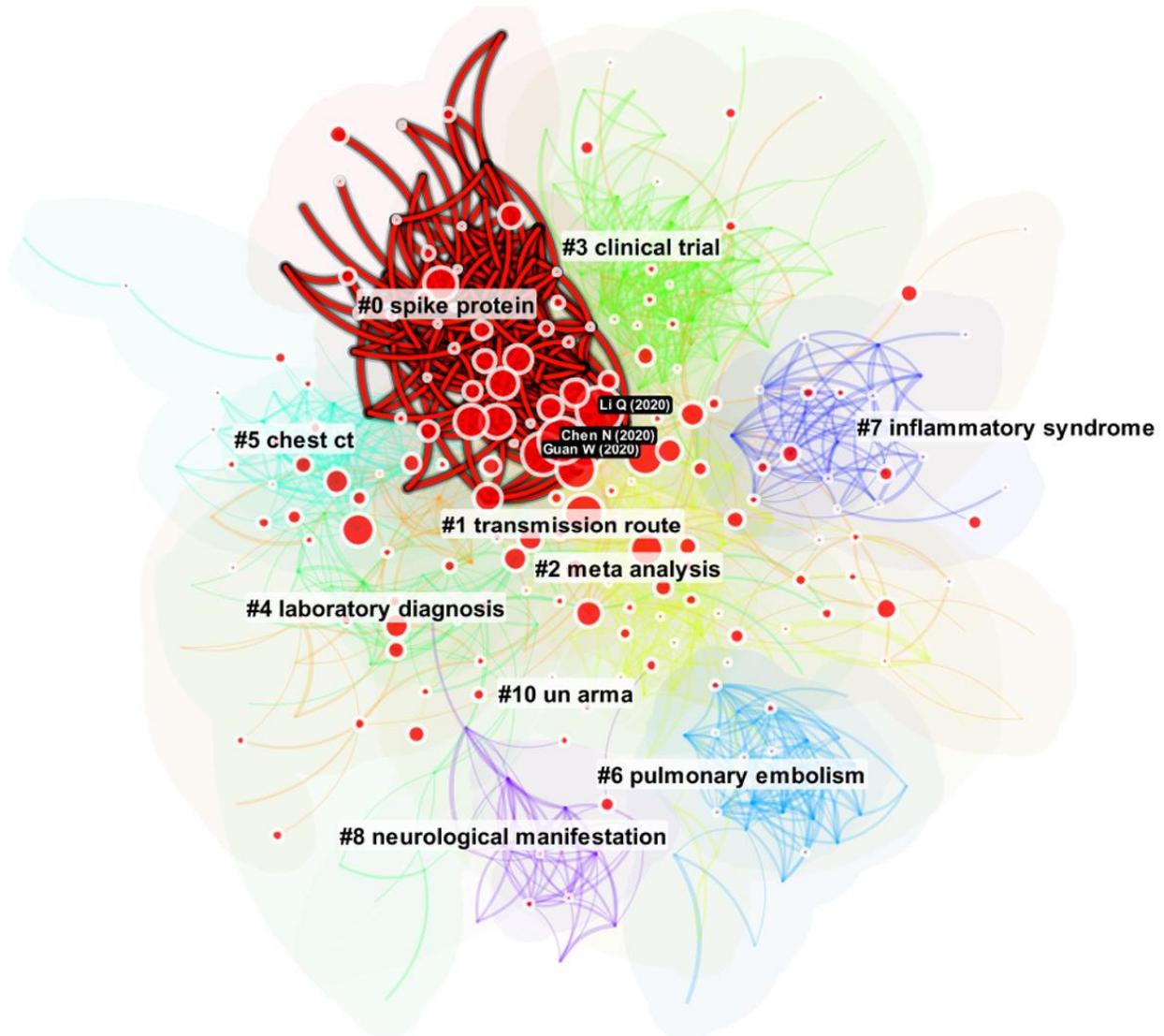

*Figure 8. References associated with the strongest sentiment of uncertainty are from Cluster #0 spike protein.*

Figure 9 shows a newly published article identified with a high transformative potential based on the centrality divergence metric, one of the three structural variation metrics. The article is also ranked high on another metric – modularity change. The visualization reveals the distribution of the references cited by the article in different clusters. The next question is why these references from different areas are cited by the same citing article.

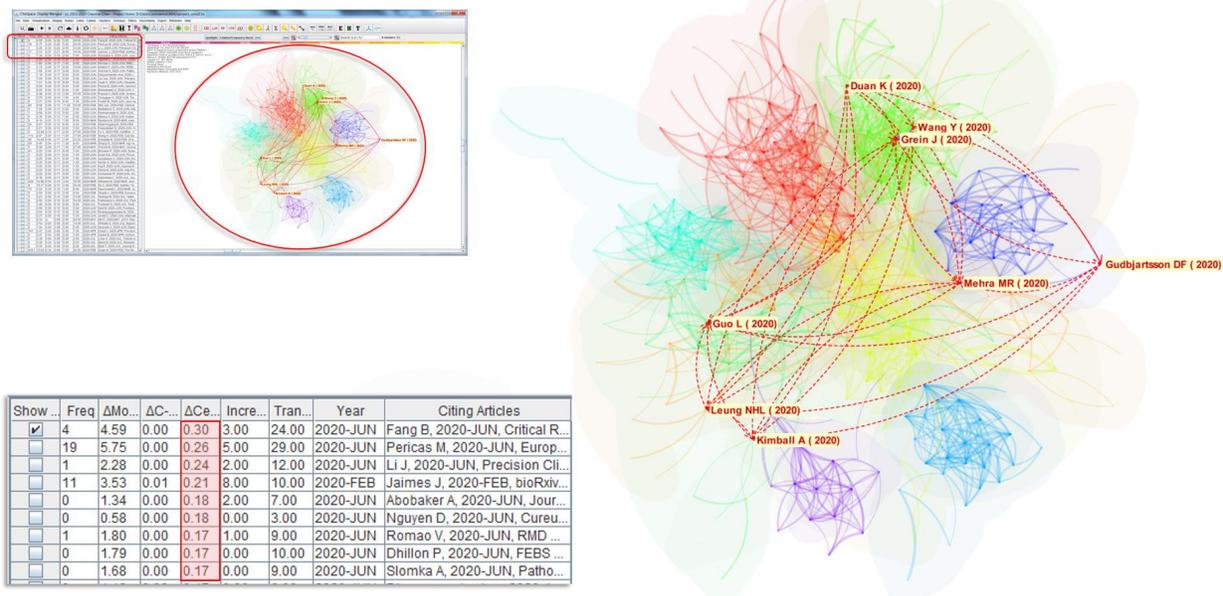

*Figure 9. An article identified by SVA with a high transformative potential according to centrality divergence.*

The review article (Fang and Meng 2020) was published in June 2020. Its citation contexts are not found on MAG at the time of writing. We manually located the citation contexts from the original publication. Table 4 shows the citation contexts of four of the references cited by (Fang and Meng 2020). Each of the references is cited in different sections on distinct topics, which partially explains the diverse footprint found by SVA.

*Table 4. Citation contexts of the references cited by (Fang and Meng 2020).*

| Cluster | Reference | Citation Context | Section Heading |
|---|---|---|---|
| #4 laboratory diagnosis | **Guo L**, Ren L, Yang S, et al. Profiling early humoral response to diagnose novel coronavirus disease (COVID-19). Clin Infect Dis. 2020. DOI:10.1093/cid/ciaa310 | Infection of SARS-CoV-2 triggers the host humoral response, leading to generation of antibodies including IgA, IgM, and IgG against SARS-CoV-2 [**78**]. | #5 Serology testing |
| | | Another study with N protein-based ELISA on 208 plasma samples from 82 confirmed and 58 probable cases revealed that the median time for detection of IgM and IgA was 5 days while for IgG it was 14 days after symptom onset [**78**]. | #5 Serology testing |
| #8 neurological manifestation | **Kimball A**, Hatfield KM, Arons M, et al.; CDC COVID-19 Investigation Team. Asymptomatic and presymptomatic SARS-CoV-2 infections in residents of a long-term care skilled nursing facility - King County, Washington, March 2020. MMWR Morb Mortal Wkly Rep. 2020;69(13):377–381. | A PCR screening test on 78 residents in a long-term care nursing home in Washington State resulted in the detection of 10 symptomatic, 10 pre-symptomatic, and 3 asymptomatic cases [**101**]. symptomatic, 10 pre-symptomatic, and 3 asymptomatic cases [**101**]. | #4 Large scale screening to detect asymptomatic or pre-symptomatic cases |

| #3 clinical trial | **Grein J**, Ohmagari N, Shin D, et al. Compassionate use of remdesivir for patients with severe Covid-19. N Engl J Med. 2020. DOI:10.1056/NEJMoa2007016 | Preliminary results showed that about 68% patients with severe COVID-19 treated with compassionate-use of remdesivir had clinical improvement [**14**], whereas a randomized, double-blind, and placebo-controlled multicenter trial showed remdesivir was not associated with significant clinical benefits [15], suggesting additional efficacy studies will be needed to demonstrated remdesivir's benefit. | #1 Introduction |
|---|---|---|---|
| #7 inflammatory syndrome | **Mehta P**, McAuley DF, Brown M, et al.; HLH Across Speciality Collaboration, UK. COVID-19: consider cytokine storm syndromes and immunosuppression. Lancet. 2020;395(10229):1033–1034. | Possible roles of cytokine storm syndrome that leads to critical disease and death of COVID-19 patients have been discussed [**54**,55]. | #3 Clinical characteristics of COVID-19 |
| | | In fact, compared with non-intensive care unit (ICU) patients, ICU patients had higher plasma levels of IL-2, IL-6, IL-7, granulocyte-colony stimulating factor, interferon-γ inducible protein 10, monocyte chemo-attractant protein 1, macrophage inflammatory protein 1-α, and TNFα [7,48,49,**54**]. | #3 Clinical characteristics of COVID-19 |
| | | In critically ill patients, cytokines and other biomarkers are significantly changed and measurement of these biochemical markers can be used to determine the severity and mortality of the disease [7,48,50,**54**]. | #3 Clinical characteristics of COVID-19 |

A practical question in constructing a systematic review of a subject is that given the current landscape of its literature, where are the promising areas to watch for future growth? Where can one expect the next breakthrough? In the case of the COVID-19 pandemic, monitoring the spread of the virus and the development of new vaccines are among the top of such a list. About 10% of the records in the MAG-based COVID-19 dataset have corresponding citation contexts available. Nearly 15% of them contained the words vaccine or vaccination. SVA enables researchers to watch activities in valleys between hills on the current literature landscape, i.e. innovative connections that link thematic islands. The literature-based discovery community pioneered by Don Swanson has been devoted it to uncover this type of connections that are missing or weak in the current literature (Chen and Song 2019).

Table 5 shows a list of articles identified with the strongest transformative potentials in terms of modularity change, which is essentially consistent with the citation-based ranking, suggesting that the validity of the list can be further verified in the near future when their citation contexts become available. Most of the articles on the list are published in June 2020, slightly over two months ago. The second one on the list has already been cited 19 times. It shows how fast the COVID-19 literature is growing as well as a promising visibility of this article. The title of the article is "COVID-19: From epidemiology to treatment." Interested readers may revisit this list regularly in the next few months and monitor whether their transformative potential is strengthened and realized or diminished and forgotten.

*Table 5. Some of the articles with the strongest transformative potentials in terms of M for modularity change. C-L is for cluster linkage. C-D is for centrality divergence.*

| MAG ID | M | C-L | C-D | Harmonic | Citations | NR | Title | Reference |
|---|---|---|---|---|---|---|---|---|
| 3006967091 | 9.082 | 0.242 | 0.134 | 0.257 | 60 | 32 | 2019 novel coronavirus of pneumonia in wuhan china emerging attack and management strategies | She Jun 2020-FEB Clinical and translational medicine V9 P1 DOI 10.1186/S40169-020-00271-Z |
| 3033364035 | 5.748 | 0.020 | 0.260 | 0.055 | 19 | 105 | covid 19 from epidemiology to treatment | Pericas M 2020-JUN European Heart Journal V41 P2092 DOI 10.1093/EURHEARTJ/EHAA462 |
| 3036326077 | 4.594 | 0.016 | 0.303 | 0.047 | 4 | 116 | the laboratory's role in combating covid 19 | Fang B 2020-JUN Critical Reviews in Clinical Laboratory Sciences V0 P1 DOI 10.1080/10408363.2020.1776675 |
| 3006282354 | 3.531 | 0.105 | 0.209 | 0.206 | 11 | 57 | structural modeling of 2019 novel coronavirus ncov spike protein reveals a proteolytically sensitive activation loop as a distinguishing feature compared to sars cov and related sars like coronaviruses | Jaimes J 2020-FEB bioRxiv V0 P DOI 10.1101/2020.02.10.942185 |
| 3035754070 | 2.984 | 0.010 | 0.138 | 0.029 | 0 | 80 | covid 19 and the cardiovascular system a review of current data summary of best practices outline of controversies and illustrative case reports | Prasad A 2020-JUN American Heart Journal V226 P174 DOI 10.1016/J.AHJ.2020.06.009 |
| 3028751559 | 2.277 | 0.008 | 0.239 | 0.024 | 1 | 123 | the epidemiology and therapeutic options for the covid 19 | Li J 2020-JUN Precision Clinical Medicine V3 P71 DOI 10.1093/PCMEDI/PBAA017 |
| 3037777580 | 1.796 | 0.006 | 0.174 | 0.018 | 1 | 56 | rheumatology practice amidst the covid 19 pandemic a pragmatic view | Romao V 2020-JUN RMD Open V6 P DOI 10.1136/RMDOPEN-2020-001314 |
| 3032928719 | 1.787 | 0.007 | 0.170 | 0.020 | 0 | 104 | covid 19 breakthroughs separating fact from fiction | Dhillon P 2020-JUN FEBS Journal V0 P DOI 10.1111/FEBS.15442 |
| 3036140173 | 1.684 | 0.006 | 0.170 | 0.018 | 0 | 134 | coronavirus disease 2019 covid 19 a short review on hematological manifestations | Slomka A 2020-JUN Pathogenetics V9 P493 DOI 10.3390/PATHOGENS9060493 |
| 3033243614 | 1.338 | 0.005 | 0.176 | 0.014 | 0 | 51 | extrapulmonary and atypical clinical presentations of covid 19 | Abobaker A 2020-JUN Journal of Medical Virology V0 P DOI 10.1002/JMV.26157 |
| 3036886562 | 1.194 | 0.004 | 0.170 | 0.012 | 0 | 29 | covid 19 and advanced practice registered nurses frontline update | Diezsampedro Ana 2020-JUN The Journal for Nurse Practitioners V0 P DOI 10.1016/J.NURPRA.2020.06.014 |
| 3039476797 | 1.039 | 0.003 | 0.150 | 0.010 | 0 | 157 | covid 19 progress in diagnostics therapy and vaccination | Liu Xue 2020-JUN Theranostics V10 P7821 DOI 10.7150/THNO.47987 |
| 3035971111 | 1.012 | 0.003 | 0.148 | 0.010 | 0 | 42 | gastrointestinal manifestations of covid 19 | Ouali S 2020-JUN Cleveland Clinic Journal of Medicine V0 P DOI 10.3949/CCJM.87A.CCC049 |

## Discussion

The illustrative examples demonstrate the advantages and further potentials of integrating MAG for the study of a rapidly growing body of literature over a monthly time scale, comparing to a much longer time period typically used in contemporary scientometric studies. Our experience shows that it is feasible for end users to construct their own datasets at a pace of their own choice. More importantly, this method opens a wide range of possibilities for researchers to compare different bibliographic databases. Such possibilities until recently have been limited to the few due to the resource-demanding nature (Chen and Song 2019, Visser, Eck et al. 2020).

At the basic level, we have demonstrated that end users would be able to conduct scientometric studies using commonly used methods such as DCA in the same way as they would with datasets from the long established but relatively more selective sources such as the Web of Science. At the same time, an equivalent topic search in the Web of Science returned 29,858 records, whereas CORD-19 has reached 130,000 records. MAG is in the same category as the Lens and Dimensions within the range of 70,000~90,000 records.

At the next level, integrating citation contexts with the visual analytic workflow significantly extends the depth of scientometric studies. The ability to cross-reference between structural indicators and linguistic cues of uncertainties has a great potential to enrich the sense-making process when facing a rapidly growing body of the literature. The integrated method enables analysts to visually inspect all the instances in which the mean incubation period are discussed with reference to a particular study. The interactive visualization considerably reduces the overhead and cognitive load. Such improvements are promising examples of integrating network-based visualization and analytic approaches with text-based visual exploration guided by different types of uncertainty metrics.

At the third level, our examples have demonstrated that the integrated approach can further push the envelope of scientometric studies and the provision of citation contexts opens up valuable opportunities for researchers to monitor and track articles identified with transformative potentials. It is also conceivable how one may extend studies along the direction demonstrated in (Bornmann, Wray et al. 2020), which resembles closed search when researchers identify concepts they would like to trace as opposed to open-ended search when researchers would welcome concepts identified by computational models and/or interactive visual exploration.

MAG has its limitations. The most noticeable one is the relatively low rate of records with citation contexts at the time of writing. We found about 10% of the records in the dataset have citation contexts. We tested the distribution of the records with citation contexts by overlaying the network of records with citation contexts on the network of all records. The result is assuring – the records with citation contexts seem to be located in the core of the larger network, whereas the missing ones tend to be located at the peripherals of the network. Even with the current level of the available citation contexts, we are able to reveal answers to some of the specific questions such as the consensus on the mean incubation period in the literature. Encouraged by the recent review of MAS (Wang, Shen et al. 2019), we believe MAS will continue to improve and the overall quality of MAG as a data source will steadily increase. The current MAS allows free API access to the MAG with a minor compromise in terms of the limit of API calls per minute and the total calls per month. MAS provides an alternative for users to host their own copy of MAG so that these restrictions can be eliminated. This is another reason why we consider MAG as a promising component to integrate in the workflow of visual analytic studies of the literature.

**Conclusion**

In conclusion, we have introduced a visual analytic method that can overcome a few significant shortcomings in the practice of conducting scientometric studies of the literature. The method provides a flexible and extensible approach that allows researchers to tailor the breadth and the depth of a dataset to meet their own requirements. Researchers are enabled to apply existing tools to such self-constructed datasets on topics of their interest at a pace of their own choice. The easy access to citation contexts provides valuable contributions to the workflow of visual analytic studies of the literature. Integrating citation contexts with network-centric analytic paradigms may serve as a stepping stone to a fully integrated workflow for the study of scientific literature at all levels of details as well as insightful patterns and trends.

**Acknowledgements**

The author would like to acknowledge the support of the Science of Science and Innovation Policy (SciSIP) Program of the National Science Foundation (Award #SMA-1633286).